\documentclass[aps,showpacs,superscriptaddress,preprint]{revtex4}%
\usepackage{amsfonts}
\usepackage{amsmath}
\usepackage{amssymb}
\usepackage{graphicx}%
\begin{document}
\title{Decoherence of a quantum system coupled to an XY spin chain: Role of the
initial state of the spin chain}
\author{Zi-Gang Yuan}
\affiliation{School of Science, Beijing University of Chemical Technology,Beijing 100029,
People's Republic of China}
\author{Ping Zhang}
\affiliation{Institute of Applied Physics and Computational Mathematics, P.O. Box 8009,
Beijing 100088, China}
\author{Shu-Shen Li}
\affiliation{State Key Laboratory for Superlattices and Microstructures, Institute of
Semiconductors, Chinese Academy of Sciences, P.O. Box 912, Beijing 100083, China}
\keywords{Decoherence, Loschmidt echo, Gaussian decay.}
\pacs{03.65.Vf, 75.10.Pq, 05.30.Pr, 42.50.Vk}

\begin{abstract}
We study the decoherence of a coupled quantum system consisting of a central
spin and its correlated environment described by a general $XY$ spin-chain
model. We make it clear that the evolution of the coherence factor sensitively
depends on the initial states of the environment spin-chain. Specially, the
dynamical evolution of the coherence factor of the central spin is numerically
and analytically investigated in both weak and strong coupling cases for
different initial states including thermal equilibrium state. In both weak and
strong coupling regimes, the decay of the coherence factor can be approximated
by a Gaussian and in the strong coupling regime the coherence factor oscillate
rapidly under a Gaussian envelope. The width of the Gaussian decay (envelope)
has been studied in details and we explained the origin of the so-called
universal regime.

\end{abstract}
\maketitle

Docoherence induced by coupling a quantum system with an environment is one of
the most important different features of opened quantum systems from isolated
ones. It refers to the process that turns the system from quantum coherent
pure states into classical mixed states. Usually, this process will destroy
the coherence between the pointer states corresponding to their eigenvalues in
a short time \cite{Zurek1} and is a major obstacle in quantum information
processing (QIP) that use the coherent entangled states as resources
\cite{Weiss,Breuer}. Hence the study of docoherence is important for
understanding quantum physics and the implementation of QIP.

The decoherence process depends on the effective Hamiltonian and the initial
state of the environment. The effects of the effective Hamiltonian on the
decoherence has been studied in many papers
\cite{Cormick1,Cormick2,You,Cheng,Lian,Quan,Yuan1,Yuan2,Liu,Cucc1,Cucc2,Ai,Li,Sun,Damski,Cincio}%
. Especially, dramatic manifestation of the decoherence has often been found
in the vicinity of the quantum critical point of the effective Hamiltonian.
Hence much work have been focused on the critical properties of the
decoherence \cite{Quan,Yuan1,Cheng,Ai,Li,Sun,Damski,Cincio}. Whereas, the
dependence of the decoherence on the initial state of the environment has been
rarely mentioned \cite{Win,Moro} and thus will be studied in this paper. In
particular, we will focus on the Gaussian decay
\cite{Cucc1,Cucc2,Zurek2,Cormick1,Yuan2,Rossini} and explain the origin of the
so-called universal regime of the Gaussian decay \cite{Cucc1,Cormick1}.

The dynamical evolution of the reduced density matrix may be used to describe
the decoherence process. For a two-level qubit system such as a central spin,
which is coupled to an environment of an $XY$ spin chain, the coefficients of
the off-diagonal terms in the reduced density matrix of the system, named as
\textquotedblleft coherence factor\textquotedblright\ in the following
discussion in this paper, may describe the degree of the decoherence. It was
found that in such a simple model and with a few additional generic
assumptions, the coherence factor displays a Gaussian decay. So the Gaussian
decay is general and important for the decoherence research. Furthermore, a
universal regime of the Gaussian envelope has been found when the coupling
strength is large enough, which means that the envelope of the decay of the
coherence factor in the system is Gaussian with a width independent of the
system-environment coupling strength \cite{Cucc1,Cormick1}. While in another
case, the Gaussian width in the strong coupling regime may be proportional to
the coupling strength \cite{Yuan2}.

Now we introduce the Hamiltonian and the model. We consider a two-level
quantum system (central spin) transversely coupled to an environment which is
described by one-dimensional $XY$ spin chain model. The total Hamiltonian is
given by $H$=$H_{E}^{\lambda}$+$H_{I}$, where (we take $\hbar$=$1$)%
\begin{align}
H_{E}^{\lambda}  &  =-\overset{N}{\sum_{l=1}}\left(  \frac{1+\gamma}{2}%
\sigma_{l}^{x}\sigma_{l+1}^{x}+\frac{1-\gamma}{2}\sigma_{l}^{y}\sigma
_{l+1}^{y}+\lambda\sigma_{l}^{z}\right)  ,\label{n1}\\
H_{I}  &  =-g\sigma^{z}\overset{N}{\sum_{l=1}}\sigma_{l}^{z}.\nonumber
\end{align}
Here $H_{E}^{\lambda}$ denotes the Hamiltonian of the environmental spin chain
and $H_{I}$ denotes the interaction between the central spin and the
environment. $\sigma^{\alpha}$ ($\alpha$=$x$, $y$, $z$) and $\sigma
_{l}^{\alpha}$ are the Pauli matrices used to describe the central spin and
the $j$th spin of the spin chain, respectively. The parameters $\lambda$
characterizes the strength of the spin interaction and the intensity of the
magnetic filed applied along the $z$ axis respectively, and $\gamma$ measures
the anisotropy in the in-plane interaction. $N$ is the total site number in
the spin chain.

Before going any further, we explain how we dress the parameter of the
intensity of the magnetic filed $\lambda$ and the Hamiltonian $H_{E}^{\lambda
}$ in this paper. To explore the dependence of the decoherence on the initial
states of the spin-chain, we should analyze various initial states. A natural
and simple choice of the initial state is the ground state of the initial
Hamiltonian $H_{E}^{\left(  \lambda_{i}\right)  }$ at time $t$=$0$ which may
be different from the evolving Hamiltonian $H_{E}^{\left(  \lambda_{e}\right)
}$ for time $t$%
$>$%
$0$. Both of the the initial Hamiltonian $H_{E}^{\left(  \lambda_{i}\right)
}$ and the evolving Hamiltonian $H_{E}^{\left(  \lambda_{e}\right)  }$ are
defined as $H_{E}^{\lambda}$ in Eq. (\ref{n1}) by replacing $\lambda$ with
$\lambda_{i}$ and $\lambda_{e}$, respectively. That is, we assume that the
coupling between the central system and the environment begin at $t$=$0$ and
there may be a sudden change for the intensity $\lambda$ of the Hamiltonian at
$t$=$0$. For simplicity, in the following of this paper we use $H$%
=$H_{E}^{\left(  \lambda_{e}\right)  }$+$H_{I}$ to represent the total
Hamiltonian for $t$%
$>$%
$0$. Furthermore, we use $\lambda_{+}$\texttt{=}$\lambda_{e}\mathtt{+}g$ and
$\lambda_{-}\mathtt{=}\lambda_{e}\mathtt{-}g$ to dress the intensity of the
magnetic filed for two effective Hamiltonians $H_{E}^{\left(  \lambda
_{+}\right)  }$ and $H_{E}^{\left(  \lambda_{-}\right)  }$, which are defined
as $H_{E}^{\lambda}$ in Eq. (\ref{n1}) by replacing $\lambda$ with
$\lambda_{+}$ and $\lambda_{-}$, respectively. So in this paper we label the
intensity of the magnetic field with four types: $\lambda_{i}$, $\lambda_{e}$,
$\lambda_{+}$ and $\lambda_{-}$.

Following Ref. \cite{Yuan2}, we can rewrite the total Hamiltonian $H$ as%
\begin{equation}
H=\left\vert 0\right\rangle \left\langle 0\right\vert \otimes H_{E}^{\left(
\lambda_{+}\right)  }+\left\vert 1\right\rangle \left\langle 1\right\vert
\otimes H_{E}^{\left(  \lambda_{-}\right)  }, \label{n2}%
\end{equation}
where $\left\vert 0\right\rangle $ and $\left\vert 1\right\rangle $ denote the
eigenstates of $\sigma^{z}$ with eigenvalues of $\pm1$. $H_{E}^{\left(
\lambda_{+}\right)  }$ and $H_{E}^{\left(  \lambda_{-}\right)  }$ are the
corresponding effective Hamiltonians of the spin chain. $H_{E}^{\left(
\lambda_{j}\right)  }$($j$=$i,e,+,-$) can be diagonalized by standard
procedure \cite{Sach}. As the first step, we define the conventional
Jordan-Wigner (JW) transformation as follows%
\begin{align}
\sigma_{l}^{x}  &  =\underset{m<l}{%
{\displaystyle\prod}
}\left(  1-2a_{m}a_{m}^{\dag}\right)  \left(  a_{l}+a_{l}^{\dag}\right)
,\label{n3a}\\
\sigma_{l}^{y}  &  =-i\underset{m<l}{%
{\displaystyle\prod}
}\left(  1-2a_{m}a_{m}^{\dag}\right)  \left(  a_{l}-a_{l}^{\dag}\right)
,\label{n3b}\\
\sigma_{l}^{z}  &  =1-2a_{l}a_{l}^{\dag}, \label{n3c}%
\end{align}
which maps spins to one-dimensional spinless fermions with creation
(annihilation) operators $a_{l}^{\dag}$ ($a_{l}$). After a straightforward
derivation, the Hamiltonians become%
\begin{equation}
H_{E}^{\left(  \lambda_{j}\right)  }=-\overset{N}{\sum_{l=1}}\left[  \left(
a_{l+1}^{\dag}a_{l}+a_{l}^{\dag}a_{l+1}\right)  +\gamma\left(  a_{l+1}%
a_{l}+a_{l}^{\dag}a_{l+1}^{\dag}\right)  -\lambda_{j}\left(  1-2a_{l}^{\dag
}a_{l}\right)  \right]  . \label{n4}%
\end{equation}
In the second step, we introduce the Fourier transformation of the fermionic
operators described by $a_{k}$=$\frac{1}{\sqrt{N}}%
{\textstyle\sum\nolimits_{l}}
a_{l}e^{-i2\pi lk/N}$, with $k$=$-M+1,..,M;M=N/2$. The Hamiltonian (\ref{n4})
can be diagonalized by transforming the fermion operators to momentum space
and then using the Bogoliubov transformation. The final result is%
\begin{align}
H_{E}^{\left(  \lambda_{j}\right)  }  &  =\overset{M}{\underset{k=1}{%
{\displaystyle\sum}
}}H_{E}^{\left(  k,\lambda_{j}\right)  }\nonumber\\
&  =\overset{M}{\underset{k=1}{%
{\displaystyle\sum}
}}\Omega_{k}^{\left(  \lambda_{j}\right)  }\left(  b_{k,\lambda_{j}}^{\dag
}b_{k,\lambda_{j}}-\frac{1}{2}\right)  , \label{n5}%
\end{align}
where $H_{E}^{\left(  k,\lambda_{j}\right)  }$=$\Omega_{k}^{\left(
\lambda_{j}\right)  }\left(  b_{k,\lambda_{j}}^{\dag}b_{k,\lambda_{j}}%
-\frac{1}{2}\right)  $ and the energy spectrum is given by%
\begin{equation}
\Omega_{k}^{\left(  \lambda_{j}\right)  }=2\sqrt{\left(  \epsilon_{k}^{\left(
\lambda_{j}\right)  }\right)  ^{2}+\gamma^{2}\sin^{2}\frac{2\pi k}{N}}
\label{n6}%
\end{equation}
with $\epsilon_{k}^{\left(  \lambda_{j}\right)  }$=$\lambda_{j}-\cos\frac{2\pi
k}{N}$, and the correspoding Bogoliubov transformed fermion operators are
defined by%
\begin{equation}
b_{k,\lambda_{j}}=\cos\frac{\theta_{k}^{\left(  \lambda_{j}\right)  }}%
{2}a_{k,\lambda_{j}}-i\sin\frac{\theta_{k}^{\left(  \lambda_{j}\right)  }}%
{2}a_{-k,\lambda_{j}}^{\dag} \label{n7}%
\end{equation}
with angles $\theta_{k}^{\left(  \lambda_{j}\right)  }$ =$\arccos
[2\epsilon_{k}^{\left(  \lambda_{j}\right)  }/\Omega_{k}^{\left(  \lambda
_{j}\right)  }].$ The corresponding ground state $\left\vert G\right\rangle
_{\lambda_{j}}$ of $H_{E}^{\left(  \lambda_{j}\right)  }$ is the vacuum of the
fermionic modes described by $b_{k,\lambda_{j}}\left\vert G\right\rangle
_{\lambda_{j}}$=$0$ for any $b_{k,\lambda_{j}}$ and can be written as
\begin{equation}
\left\vert G\right\rangle _{\lambda_{j}}=\prod\nolimits_{k=1}^{M}\left\vert
G\right\rangle _{\lambda_{j}}^{k}=%
{\textstyle\prod\nolimits_{k=1}^{M}}
\left[  \cos\frac{\theta_{k}^{\left(  \lambda_{j}\right)  }}{2}\left\vert
0\right\rangle _{k}\left\vert 0\right\rangle _{-k}+i\sin\frac{\theta
_{k}^{\left(  \lambda_{j}\right)  }}{2}\left\vert 1\right\rangle
_{k}\left\vert 1\right\rangle _{-k}\right]  , \label{n7a}%
\end{equation}
where%
\begin{equation}
\left\vert G\right\rangle _{\lambda_{j}}^{k}=\cos\frac{\theta_{k}^{\left(
\lambda_{j}\right)  }}{2}\left\vert 0\right\rangle _{k}\left\vert
0\right\rangle _{-k}+i\sin\frac{\theta_{k}^{\left(  \lambda_{j}\right)  }}%
{2}\left\vert 1\right\rangle _{k}\left\vert 1\right\rangle _{-k}, \label{n7b}%
\end{equation}
$\left\vert 0\right\rangle _{k}$ and $\left\vert 1\right\rangle _{k}$ denote
the vacuum and single excitation of the $k$th mode $a_{k,\lambda_{j}}$,
respectively. It is straightforward to see that the operator $b_{k,\lambda
_{j}}$ is related to the operator $b_{k,\lambda_{j^{\prime}}}$ by the
following relation:%
\begin{equation}
b_{k,\lambda_{j}}=\left[  \cos\alpha_{k}^{\left(  \lambda_{jj^{\prime}%
}\right)  }\right]  b_{k,\lambda_{j^{\prime}}}-i\left[  \sin\alpha
_{k}^{\left(  \lambda_{jj^{\prime}}\right)  }\right]  b_{-k,\lambda
_{j^{\prime}}}^{\dag}, \label{n8}%
\end{equation}
where $j$=$i,e,+,-$, and $\alpha_{k}^{\left(  \lambda_{jj^{\prime}}\right)  }%
$=$\frac{\theta_{k}^{\left(  \lambda_{j}\right)  }-\theta_{k}^{\left(
\lambda_{j^{\prime}}\right)  }}{2}$. As a result, the corresponding ground
states satisfy the relation%
\begin{equation}
|G\rangle_{\lambda_{j}}=\prod\nolimits_{k=1}^{M}\left[  \cos\alpha
_{k}^{\left(  \lambda_{jj^{\prime}}\right)  }+i\sin\alpha_{k}^{\left(
\lambda_{jj^{\prime}}\right)  }b_{k,\lambda_{j^{\prime}}}^{\dag}%
b_{-k,\lambda_{j^{\prime}}}^{\dag}\right]  |G\rangle_{\lambda_{j^{\prime}}}.
\label{n9}%
\end{equation}

Suppose the initial state of the total system is described by the density
matrix:%
\begin{equation}
\rho_{\mathbf{tot}}\left(  0\right)  =\rho_{s}\left(  0\right)  \otimes
\rho_{E}\left(  0\right)  , \label{n11}%
\end{equation}
where $\rho_{s}\left(  0\right)  $ and $\rho_{E}\left(  0\right)  $ is the
initial density matrix of the central system and the environment respectively.
The evolved density matrix of the total system for $t>0$ is
\begin{equation}
\rho_{\mathbf{tot}}\left(  t\right)  =U\left(  t\right)  \rho_{\mathbf{tot}%
}\left(  0\right)  U^{\dag}\left(  t\right)  , \label{n23}%
\end{equation}
where $U\left(  t\right)  $ is the time evolution matrix which can be obtained
by solving the equation%
\[
i\dot{U}\left(  t\right)  =HU.
\]
Equation (\ref{n23}) has an exact solution for a time-dependent step function
form for the magnetic field $\lambda\left(  t\right)  =\lambda_{i}+\left(
\lambda_{e}-\lambda_{i}\right)  \theta\left(  t\right)  $ which we adopt in
this work \cite{Sadiek,Alkur}. Here $\theta\left(  t\right)  $ is the usual
mathematical step function. The solution to the time evolution operator for
the Hamiltonian of the time-dependent step function form is exactly the same
as that for the Hamiltonian of $H$ for $t>0$. That is, the time evolution
operator can be expressed as
\begin{equation}
U\left(  t\right)  =\left\vert 0\right\rangle \left\langle 0\right\vert
\otimes U_{E}^{\left(  \lambda_{+}\right)  }\left(  t\right)  +\left\vert
1\right\rangle \left\langle 1\right\vert \otimes U_{E}^{\left(  \lambda
_{-}\right)  }\left(  t\right)  , \label{n10}%
\end{equation}
where $U_{E}^{\left(  \lambda_{\pm}\right)  }\left(  t\right)  $=$\exp\left[
-iH_{E}^{\left(  \lambda_{\pm}\right)  }t\right]  $ is the effective time
evolution operator dressed by $\lambda_{\pm}$. As a result, the reduced
density matrix of the central system is
\begin{align}
\rho_{S}\left(  t\right)   &  =\mathbf{Tr}_{E}\left[  \rho_{\mathbf{tot}%
}\left(  t\right)  \right] \nonumber\\
&  =\left(
\begin{array}
[c]{cc}%
\left[  \rho_{S}\left(  0\right)  \right]  _{11} & \left[  \rho_{S}\left(
0\right)  \right]  _{12}\left[  \mathbf{Tr}_{E}U_{E}^{\left(  \lambda
_{+}\right)  }\left(  t\right)  \rho_{E}\left(  0\right)  U_{E}^{\dag\left(
\lambda_{-}\right)  }\left(  t\right)  \right] \\
\left[  \rho_{S}\left(  0\right)  \right]  _{21}\left[  \mathbf{Tr}_{E}%
U_{E}^{\left(  \lambda_{-}\right)  }\left(  t\right)  \rho_{E}\left(
0\right)  U_{E}^{\dag\left(  \lambda_{+}\right)  }\left(  t\right)  \right]  &
\left[  \rho_{S}\left(  0\right)  \right]  _{22}%
\end{array}
\right)  . \label{n12}%
\end{align}
It reveals in Eq. (\ref{n12}) that the environmental spin chain only modulates
the off-diagonal terms of $\rho\left(  t\right)  $ through the coherence
factor\
\begin{equation}
F\left(  t\right)  =\left\vert \mathbf{Tr}_{E}\left[  U_{E}^{\left(
\lambda_{+}\right)  }\left(  t\right)  \rho_{E}\left(  0\right)  U_{E}%
^{\dag\left(  \lambda_{-}\right)  }\left(  t\right)  \right]  \right\vert ,
\label{n13a}%
\end{equation}
Equation (\ref{n13a}) is our starting point of the following derivation and
discussions. If the initial density matrix of the environmental spin chain
$\rho_{E}\left(  0\right)  $ can be factored as $\rho_{E}\left(  0\right)
$=$\prod\nolimits_{k=1}^{M}\otimes\rho_{E}^{k}\left(  0\right)  $, then%
\begin{equation}
F\left(  t\right)  =\prod\nolimits_{k=1}^{M}F_{k}\left(  t\right)  ,
\label{n14}%
\end{equation}
where
\begin{equation}
F_{k}\left(  t\right)  =\left\vert \mathbf{Tr}_{E}\left\{  \exp\left[
iH_{E}^{\left(  k,\lambda_{+}\right)  }t\right]  \rho_{E}^{k}\left(  0\right)
\exp\left[  -iH_{E}^{\left(  k,\lambda_{-}\right)  }t\right]  \right\}
\right\vert . \label{n15}%
\end{equation}

Following Refs. \cite{Sadiek,Alkur}, we can also rewrite the Hamiltonian
$H_{E}^{\left(  k,\lambda_{j}\right)  }$ and the effective time evolution
operator $U_{E}^{\left(  k,\lambda_{j}\right)  }\left(  t\right)  $%
=$\exp\left[  -iH_{E}^{\left(  k,\lambda_{j}\right)  }t\right]  $ in the basis
of $\left\vert \phi_{k}\right\rangle _{1}\mathtt{=}\left\vert 0\right\rangle
$, $\left\vert \phi_{k}\right\rangle _{2}\mathtt{=}a_{k,\lambda_{j}}^{\dag
}a_{-k,\lambda_{j}}^{\dag}\left\vert 0\right\rangle $, $\left\vert \phi
_{k}\right\rangle _{3}\mathtt{=}a_{k,\lambda_{j}}^{\dag}\left\vert
0\right\rangle $, and $\left\vert \phi_{k}\right\rangle _{4}=a_{-k,\lambda
_{j}}^{\dag}\left\vert 0\right\rangle $ (or equivelently $\left\vert
0\right\rangle _{k}\left\vert 0\right\rangle _{-k}$, $\left\vert
1\right\rangle _{k}\left\vert 1\right\rangle _{-k}$, $\left\vert
1\right\rangle _{k}\left\vert 0\right\rangle _{-k}$, and $\left\vert
0\right\rangle _{k}\left\vert 1\right\rangle _{-k}$) as follows:%
\begin{equation}
H_{E}^{\left(  k,\lambda_{j}\right)  }=\left\{
\begin{array}
[c]{cccc}%
-\Omega_{k}^{\left(  \lambda_{j}\right)  }\cos\theta_{k}^{j}-2\cos\frac{2\pi
k}{N} & i\Omega_{k}^{\left(  \lambda_{j}\right)  }\sin\theta_{k}^{j} & 0 & 0\\
-i\Omega_{k}^{\left(  \lambda_{j}\right)  }\sin\theta_{k}^{j} & \Omega
_{k}^{\left(  \lambda_{j}\right)  }\cos\theta_{k}^{j}-2\cos\frac{2\pi k}{N} &
0 & 0\\
0 & 0 & -2\cos\frac{2\pi k}{N} & 0\\
0 & 0 & 0 & -2\cos\frac{2\pi k}{N}%
\end{array}
\right\}  , \label{n24}%
\end{equation}%
\begin{equation}
U_{E}^{\left(  k,\lambda_{j}\right)  }\left(  t\right)  =e^{2it\cos\phi_{k}%
}\left\{
\begin{array}
[c]{cccc}%
i\cos\theta_{k}^{j}\sin\left(  2t\Lambda_{k}^{j}\right)  +\cos\left(
2t\Lambda_{k}^{j}\right)  & \sin\theta_{k}^{j}\sin\left(  2t\Lambda_{k}%
^{j}\right)  & 0 & 0\\
-\sin\theta_{k}^{j}\sin\left(  2t\Lambda_{k}^{j}\right)  & -i\cos\theta
_{k}^{j}\sin\left(  2t\Lambda_{k}^{j}\right)  +\cos\left(  2t\Lambda_{k}%
^{j}\right)  & 0 & 0\\
0 & 0 & 1 & 0\\
0 & 0 & 0 & 1
\end{array}
\right\}  . \label{n25}%
\end{equation}
Then for specific initial state of the environmental system $\rho_{E}\left(
0\right)  $, we can derive the corresponding coherence factor $F\left(
t\right)  $.

\begin{figure}[ptb]
\begin{center}
\includegraphics[width=0.6\linewidth]{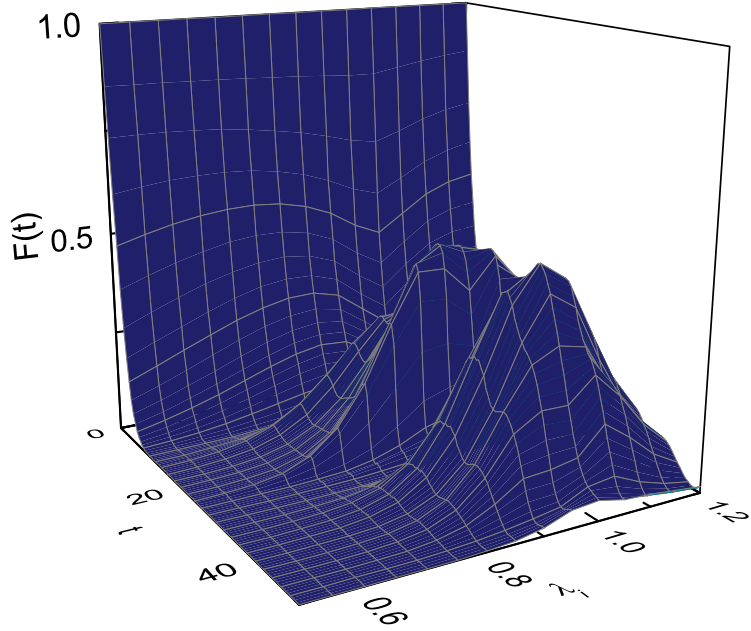}
\end{center}
\caption{(Color online). The coherence factor $F\left(  t\right)  $ as a
function of time $t$ and $\lambda_{i}$ for Ising ($\gamma$=$1$) spin chain
with a size of $N$=$100$.}%
\label{fig1}%
\end{figure}

A natural choice of the initial state of the environmental system is the
ground state $\left\vert G\right\rangle _{\lambda_{i}}$ of the initial
Hamiltonian $H_{E}^{\left(  k,\lambda_{i}\right)  }$,%
\begin{equation}
\rho_{k}\left(  0\right)  =S^{\dag}\left(
\begin{array}
[c]{cccc}%
1 & 0 & 0 & 0\\
0 & 0 & 0 & 0\\
0 & 0 & 0 & 0\\
0 & 0 & 0 & 0
\end{array}
\right)  S=\left(
\begin{array}
[c]{cccc}%
\frac{1}{2}+\frac{1}{2}\cos\theta_{k}^{i} & -\frac{i}{2}\sin\theta_{k}^{i} &
0 & 0\\
\frac{i}{2}\sin\theta_{k}^{i} & \frac{1}{2}-\frac{1}{2}\cos\theta_{k}^{i} &
0 & 0\\
0 & 0 & 0 & 0\\
0 & 0 & 0 & 0
\end{array}
\right)  , \label{n26}%
\end{equation}
where
\[
S=\left(
\begin{array}
[c]{cccc}%
\cos\frac{\theta_{k}^{i}}{2} & -i\sin\frac{\theta_{k}^{i}}{2} & 0 & 0\\
-i\sin\frac{\theta_{k}^{i}}{2} & \cos\frac{\theta_{k}^{i}}{2} & 0 & 0\\
0 & 0 & 1 & 0\\
0 & 0 & 0 & 1
\end{array}
\right)
\]
is the transformation matrix. Then we can get the coherence factor $F\left(
t\right)  $ with the Eq. (\ref{n15}). The result is%
\begin{align}
F\left(  t\right)   &  =\prod\nolimits_{k=1}^{M}|[\cos2\alpha_{k}^{\left(
\lambda_{+-}\right)  }\sin\left(  \Omega_{k}^{\left(  \lambda_{+}\right)
}t\right)  \sin\left(  \Omega_{k}^{\left(  \lambda_{-}\right)  }t\right)
+\cos\left(  \Omega_{k}^{\left(  \lambda_{+}\right)  }t\right)  \cos\left(
\Omega_{k}^{\left(  \lambda_{-}\right)  }t\right) \nonumber\\
&  +i\cos2\alpha_{k}^{\left(  \lambda_{+i}\right)  }\sin\left(  \Omega
_{k}^{\left(  \lambda_{+}\right)  }t\right)  \cos\left(  \Omega_{k}^{\left(
\lambda_{-}\right)  }t\right)  -i\cos2\alpha_{k}^{\left(  \lambda_{-i}\right)
}\sin\left(  \Omega_{k}^{\left(  \lambda_{+}\right)  }t\right)  \cos\left(
\Omega_{k}^{\left(  \lambda_{-}\right)  }t\right)  ]|, \label{n17}%
\end{align}
or equivalently
\begin{align}
F\left(  t\right)   &  =\prod\nolimits_{k=1}^{M}|-\exp\left[  it\left(
\Omega_{k}^{\left(  \lambda_{+}\right)  }+\Omega_{k}^{\left(  \lambda
_{-}\right)  }\right)  \right]  \sin\alpha_{k}^{\left(  \lambda_{+-}\right)
}\cos\alpha_{k}^{\left(  \lambda_{+i}\right)  }\sin\alpha_{k}^{\left(
\lambda_{-i}\right)  }\nonumber\\
&  +\exp\left[  it\left(  -\Omega_{k}^{\left(  \lambda_{+}\right)  }%
+\Omega_{k}^{\left(  \lambda_{-}\right)  }\right)  \right]  \cos\alpha
_{k}^{\left(  \lambda_{+-}\right)  }\sin\alpha_{k}^{\left(  \lambda
_{+i}\right)  }\sin\alpha_{k}^{\left(  \lambda_{-i}\right)  }\nonumber\\
&  +\exp\left[  it\left(  \Omega_{k}^{\left(  \lambda_{+}\right)  }-\Omega
_{k}^{\left(  \lambda_{-}\right)  }\right)  \right]  \cos\alpha_{k}^{\left(
\lambda_{+-}\right)  }\cos\alpha_{k}^{\left(  \lambda_{+i}\right)  }\cos
\alpha_{k}^{\left(  \lambda_{-i}\right)  }\nonumber\\
&  +\exp\left[  it\left(  -\Omega_{k}^{\left(  \lambda_{+}\right)  }%
-\Omega_{k}^{\left(  \lambda_{-}\right)  }\right)  \right]  \sin\alpha
_{k}^{\left(  \lambda_{+-}\right)  }\sin\alpha_{k}^{\left(  \lambda
_{+i}\right)  }\cos\alpha_{k}^{\left(  \lambda_{-i}\right)  }|. \label{n18}%
\end{align}
In Refs. \cite{Quan,Yuan1,Yuan2}, it has been shown that the coherence factor
$F\left(  t\right)  $ decay more dramatically and rapidly in the vicinity of
the quantum critical point $\lambda_{i}$=$\lambda_{e}$=$1$.$0$
\cite{Quan,Yuan1,Yuan2} for small $N$ ($N\mathtt{\approx}10^{2}$). In this
paper we will study the decoherence with different initial states. Hence in
Fig. \ref{fig1} we show dynamical evolution of the coherence facotor $F\left(
t\right)  $ as a function of time $t$ and $\lambda_{i}$ and keep $\lambda_{e}%
$=$1$ here and in the following unless specified. The other parameters are
$\gamma$=$1$, $g$=$0.05$, $N$=$10^{2}$. One can see that $F\left(  t\right)  $
with any $\lambda_{i}$ decays from unity to zero in a short time. And around
$\lambda_{i}$=$1$, there are significant revivals. This indicates that the
decoherence process with different $\lambda_{i}$ and fixed $\lambda_{e}$ may
include new interesting behavior and lead to more results. 

\begin{figure}[ptb]
\begin{center}
\includegraphics[width=0.6\linewidth]{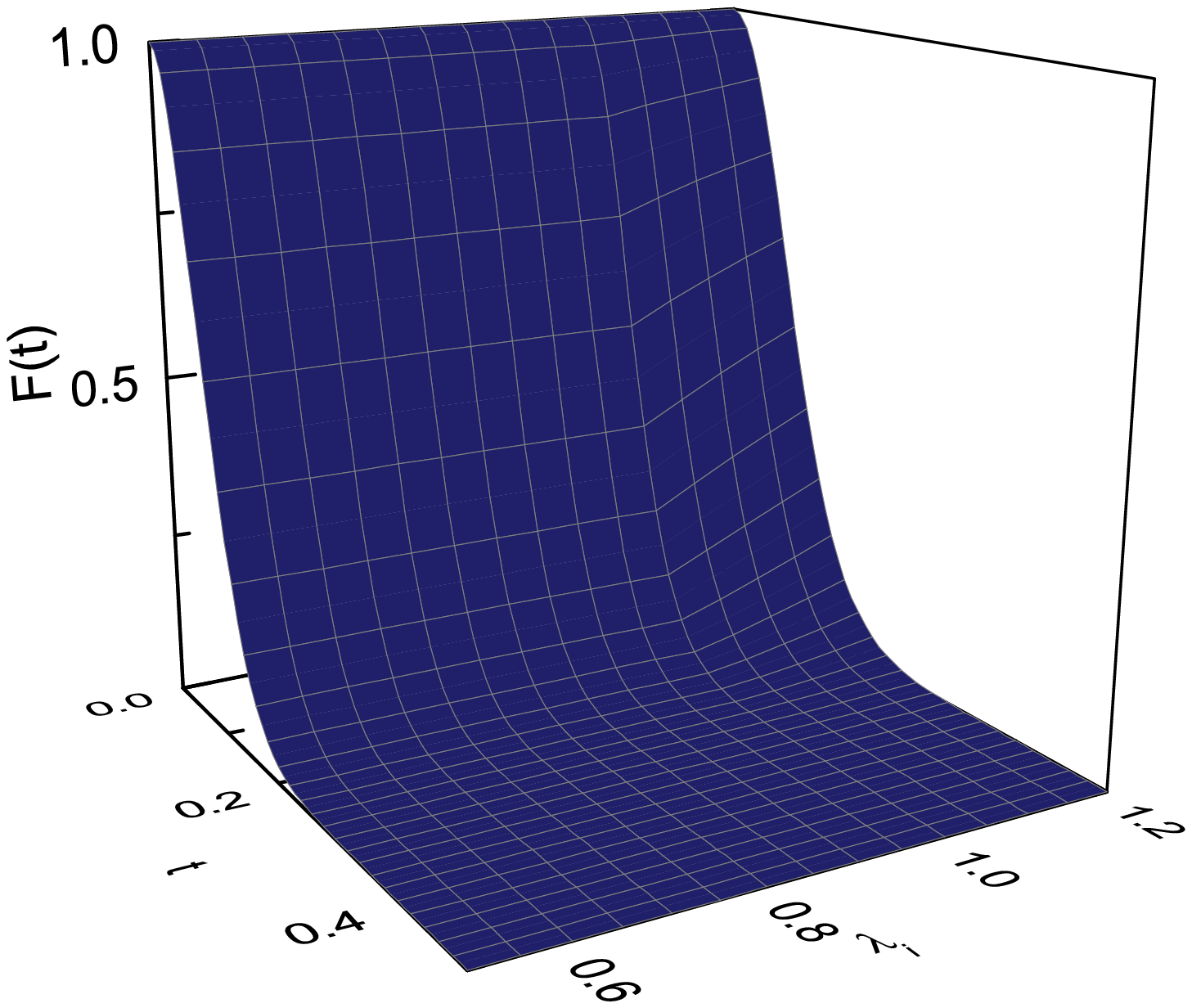}
\end{center}
\caption{(Color online). The coherence factor $F\left(  t\right)  $ as a
function of time $t$ and $\lambda_{i}$ for Ising ($\gamma$=$1$) spin chain
with a size of $N$=$1\times10^{4}$.}%
\label{fig2}%
\end{figure}

In fact, the revivals only appear for small $N$. For large $N$ which may be
used to simulate the thermodynamic limit, there are no revivals. In Fig.
\ref{fig2} we show the evolution of $F\left(  t\right)  $ as a function of
time $t$ and $\lambda_{i}$. The other parameters are $\gamma$=$1$, $g$=$0.05$,
and $N$=$10^{4}$. One can see that $F\left(  t\right)  $ decays rapidly with
no revivals for any $\lambda_{i}$. The disappearance of the revivals for large
$N$ can be understood in the following way. Notice that $F\left(  t\right)  $
is a product of a series of $k$ modes. At $t$=$0$, the module of every mode is
unity. After that, some of them decays remarkably, and the others remain unity.
The rapid decay is induced by the effect of QPT of the evolving Hamiltonian
$\lambda_{e}$=$1$. The effect of QPT come down to the disappearance of the
energy gap between the ground state and the excited states. For small $N$, the
number of the modes whose energy gaps nearly disappear is small. As a result,
the evolution of the coherence factor $F\left(  t\right)  $ depends on the
evolution of $F_{k}\left(  t\right)  $ of few modes. Hence there are revivals.
For large $N$, the number of the modes whose energy gaps nearly disappear is
large and thus leads to a chaos result with no revivals.

In fact, the decay of the coherence factor $F\left(  t\right)  $ is
Gaussian. The Gaussian decay arises from the following expression
\begin{align}
r(t)  &  =\overset{M}{\underset{k}{%
{\displaystyle\prod}
}}r_{k}(t)\label{n27a}\\
&  =\overset{M}{\underset{k}{%
{\displaystyle\prod}
}}\left[  \left\vert \alpha_{k}\right\vert ^{2}\exp\left(  i\Omega
_{k}t\right)  +\left\vert \beta_{k}\right\vert ^{2}\exp\left(  -i\Omega
_{k}t\right)  \right]  , \label{n27b}%
\end{align}
where the coefficients $\left\vert \alpha_{k}\right\vert ^{2}$ and $\left\vert
\beta_{k}\right\vert ^{2}$ satisfy $\left\vert \alpha_{k}\right\vert ^{2}%
$+$\left\vert \beta_{k}\right\vert ^{2}$=$1$. The value of such an expression
can be understood as the \textquotedblleft random walk\textquotedblright%
\ problem discussed in Refs. \cite{Cucc1,Cucc2}, in which the authors
considered the distribution of $M$-step random walk. Each step correlates to a
random variable taking the value $+\Omega_{k}$ or $-\Omega_{k}$ with
probability $\left\vert \alpha_{k}\right\vert ^{2}$ or $\left\vert \beta
_{k}\right\vert ^{2}$ respectively. We denote $a_{k}$ and $b_{k}$ the mean
value and its variance of the random variable. If Lindeberg condition, which
demands that the cumulative variances is finite, is satisfied, this
distribution of energies yields a approximately Gaussian time dependence of
$r(t)$ as%
\begin{equation}
|r(t)|\approx\exp\left(  -\frac{s_{M}^{2}t^{2}}{2}\right)  , \label{n28}%
\end{equation}
where
\begin{equation}
s_{M}^{2}=\underset{k}{\overset{M}{%
{\displaystyle\sum}
}}b_{k}^{2} \label{n29}%
\end{equation}
is the cumulative variance.

\begin{figure}[ptb]
\begin{center}
\includegraphics[width=0.8\linewidth]{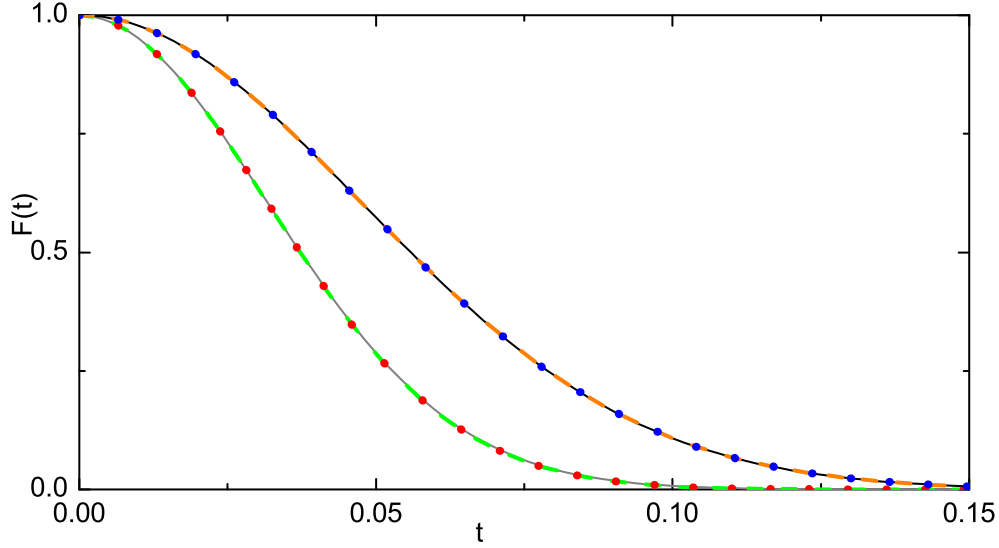}
\end{center}
\caption{(Color online). The evolution of the coherence factor $F\left(
t\right)  $ and its approximate analytic expressions Eqs. (\ref{n31b}) and (\ref{n33}) as
a function of time $t$ with $\lambda_{i}$=$0.5$ and
$\lambda_{i}$=$1.5$ for Ising ($\gamma$=$1$) spin chain with a size of
$N$=$10^{5}$. Here, the solid lines are drawn with Eq. (\ref{n18}), the dash
lines are drawn with Eq. (\ref{n31b}), and the dot lines are drawn with Eq.
(\ref{n33}).}%
\label{fig3}%
\end{figure}

In the expression of the coherence factor for $\left\vert
G\right\rangle _{\lambda_{i}}$, Eq. (\ref{n18}), $F_{k}$ has four terms. While
the sum of the coefficients of the four terms is still unity. This enlightens
us to consider $M$-step random walk with random varible taking the values
$\Omega_{k1}$=$+\left(  \Omega_{k}^{\left(  \lambda_{+}\right)  }+\Omega
_{k}^{\left(  \lambda_{-}\right)  }\right)  $, $\Omega_{k2}$=$-\left(
\Omega_{k}^{\left(  \lambda_{+}\right)  }+\Omega_{k}^{\left(  \lambda
_{-}\right)  }\right)  $, $\Omega_{k3}$=$+\left(  \Omega_{k}^{\left(
\lambda_{+}\right)  }-\Omega_{k}^{\left(  \lambda_{-}\right)  }\right)  $, and
$\Omega_{k4}$=$-\left(  \Omega_{k}^{\left(  \lambda_{+}\right)  }-\Omega
_{k}^{\left(  \lambda_{-}\right)  }\right)  $ with \textquotedblleft
probabilities\textquotedblright\ $p_{k1}$=$\sin\alpha_{k}^{\left(
\lambda_{+-}\right)  }\cos\alpha_{k}^{\left(  \lambda_{+i}\right)  }\sin
\alpha_{k}^{\left(  \lambda_{-i}\right)  }$, $p_{k2}$=$\cos\alpha_{k}^{\left(
\lambda_{+-}\right)  }\sin\alpha_{k}^{\left(  \lambda_{+i}\right)  }\sin
\alpha_{k}^{\left(  \lambda_{-i}\right)  }$, $p_{k3}$=$\cos\alpha_{k}^{\left(
\lambda_{+-}\right)  }\cos\alpha_{k}^{\left(  \lambda_{+i}\right)  }\cos
\alpha_{k}^{\left(  \lambda_{-i}\right)  }$, and $p_{k4}$=$\sin\alpha
_{k}^{\left(  \lambda_{+-}\right)  }\sin\alpha_{k}^{\left(  \lambda
_{+i}\right)  }\cos\alpha_{k}^{\left(  \lambda_{-i}\right)  }$, respectively.
Although the \textquotedblleft probability\textquotedblright\ may be negative,
the derivation process is similar with that given in Refs. \cite{Cucc1,Cucc2}.
For this purpose we define $a_{k}$=$\bar{\Omega}_{kl}$=$%
{\textstyle\sum\nolimits_{l=1}^{4}}
p_{kl}\Omega_{kl}$ and $b_{k}$=$%
{\textstyle\sum\nolimits_{l=1}^{4}}
\left(  p_{kl}\Omega_{kl}^{2}-p_{kl}a_{k}^{2}\right)  $ as mean value and
variance of the four random variables, respectively. After a straightforward
derivation, we get, in the case of the initial state being $\left\vert
G\right\rangle _{\lambda_{i}}$,
\begin{align}
a_{k}  &  =4g\cos\theta_{k}^{i},\label{n30a}\\
s_{M}^{2}  &  =\underset{k}{\overset{M}{%
{\displaystyle\sum}
}}b_{k}^{2}=\underset{k}{\sum}\underset{l}{\sum}\left[  p_{kl}\left(
\Omega_{kl}-a_{k}\right)  ^{2}\right]  =16g^{2}\underset{k}{\sum}\sin
^{2}\theta_{k}^{i}. \label{n30b}%
\end{align}
For large $N$, the Lindeberg condition is satisfied and the decoherence factor
is
\begin{equation}
F(t)\approx\exp\left(  -s_{M}^{2}t^{2}/2\right)  \approx\exp\left(
-8g^{2}t^{2}\underset{k}{\sum}\sin^{2}\theta_{k}^{i}\right)  . \label{n31b}%
\end{equation}
By noting that for $\gamma$=$1$,
\begin{equation}
\underset{k}{\sum}\sin^{2}\theta_{k}^{i}\approx\left\{
\begin{array}
[c]{c}%
\frac{M}{2\lambda_{i}^{2}}\\
\frac{M}{2}%
\end{array}
\right.
\begin{array}
[c]{c}%
\lambda_{i}^{2}>1\\
\lambda_{i}^{2}\leq1
\end{array}
, \label{n31}%
\end{equation}
we get%
\begin{equation}
s_{M}^{2}\approx\left\{
\begin{array}
[c]{c}%
\frac{8g^{2}M}{\lambda_{i}^{2}}\\
8g^{2}M
\end{array}
\right.
\begin{array}
[c]{c}%
\lambda_{i}^{2}>1\\
\lambda_{i}^{2}\leq1
\end{array}
. \label{n32}%
\end{equation}
Thus, the evolution of coherence factor $F(t)$ of Ising model ($\gamma$=$1$)
can be approximated by a very simple formula, \qquad%
\begin{equation}
F\left(  t\right)  \approx\left\{
\begin{array}
[c]{c}%
\exp\left(  -4Mg^{2}t^{2}/\lambda_{i}^{2}\right) \\
\exp\left(  -4Mg^{2}t^{2}\right)
\end{array}
\right.
\begin{array}
[c]{c}%
\lambda_{i}^{2}>1\\
\lambda_{i}^{2}\leq1
\end{array}
, \label{n33}%
\end{equation}
which is only affected by the coupling strength $g$, $M$ and
$\lambda_{i}$. With certain $\lambda_{i}$, bigger chain number and stronger
coupling strength, $F(t)$ will decay more rapidly. In the previous work
\cite{Cormick1,Mukh}, the relation between the width of the Gaussian decay and
the parameters of the Hamiltonian has been studied. Here we first give the
expression for the reciprocal of the width $s_{M}^{2}$ with the parameters of
the Hamiltonian for Ising ($\gamma$=$1$) model. In Fig. \ref{fig3} we show the
numerical results of $F(t)$ as a function of $t$ with $\lambda_{i}$=$0.5$ and
$\lambda_{i}$=$1.5$ respectively. These results are calculated with Eqs.
(\ref{n18}), (\ref{n31b}), and (\ref{n33}) respectively. The solid lines are
drawn with Eq. (\ref{n18}); the dash lines are drawn with Eq. (\ref{n31b});
the dot lines are drawn with E.q. (\ref{n33}). The three upper lines are drawn
with $\lambda_{i}$=$0.5$ and the lower lines are drawn with $\lambda_{i}%
$=$1.5$. One can see that the lines from either Eq. (\ref{n31b}) or
(\ref{n33}) fit the lines from Eq. (\ref{n18}) very well. From Eq. (\ref{n32})
one can also see that the critical point ($\lambda_{i}$=$1$) of the initial
Hamiltonian still plays a special role. At this critical point, the decaying
speed is not particularly high, while the derivative of $s_{M}^{2}$ with
$\lambda_{i}^{2}$ becomes discontinuous. 

\begin{figure}[ptb]
\begin{center}
\includegraphics[width=0.6\linewidth]{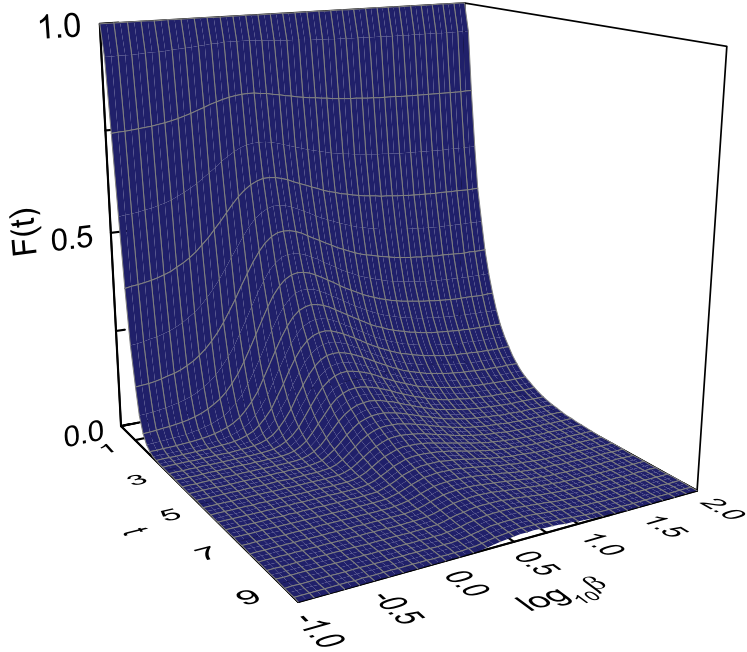}
\end{center}
\caption{(Color online). The coherence factor $F\left(  t\right)  $ as a
function of time $t$ and temperature $T$ for Ising ($\gamma$=$1$) spin chain
with a size of $N$=$200$.}%
\label{fig4}%
\end{figure}

\begin{figure}[ptbptb]
\begin{center}
\includegraphics[width=0.8\linewidth]{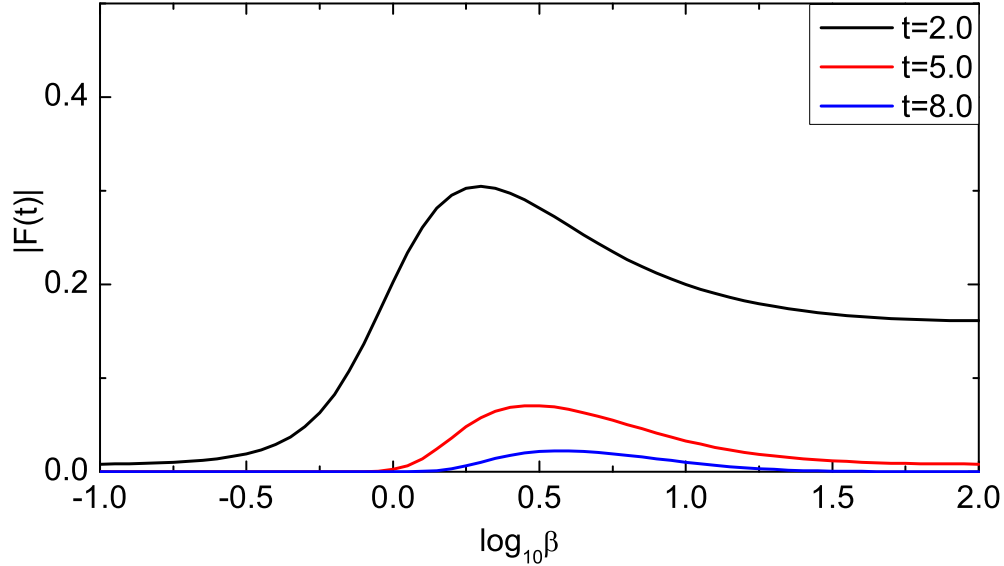}
\end{center}
\caption{(Color online). The coherence factor $F\left(  t\right)  $ as a
function of temperature $T$ at certain time for Ising ($\gamma$=$1$)
spin chain with a size of $N$=$200$. These lines are from the cross-section of Fig
\ref{fig4} at different time $t$.}%
\label{fig5}%
\end{figure}

Next, we choose thermal equilibrium state as initial state. By noting
that the partition function of the thermal equilibrium state should be
determined by the initial Hamiltonian $H_{E}^{\left(  \lambda_{i}\right)  }$,
but not $H_{E}^{\left(  \lambda_{e}\right)  }$, we can derive the expression
for the coherence factor as%
\begin{align}
F\left(  t\right)   &  =\frac{1}{Z}|[\cos2\alpha_{k}^{\left(  \lambda
_{+-}\right)  }\sin\left(  \Omega_{k}^{\left(  \lambda_{+}\right)  }t\right)
\sin\left(  \Omega_{k}^{\left(  \lambda_{-}\right)  }t\right)  +\cos\left(
\Omega_{k}^{\left(  \lambda_{+}\right)  }t\right)  \cos\left(  \Omega
_{k}^{\left(  \lambda_{-}\right)  }t\right)  ]\nonumber\\
&  \times\left[  \exp\left(  -2\beta\Omega_{k}^{\left(  \lambda_{i}\right)
}\right)  +1\right]  -i[\cos2\alpha_{k}^{\left(  \lambda_{+i}\right)  }%
\sin\left(  \Omega_{k}^{\left(  \lambda_{+}\right)  }t\right)  \cos\left(
\Omega_{k}^{\left(  \lambda_{-}\right)  }t\right) \nonumber\\
&  -\cos2\alpha_{k}^{\left(  \lambda_{-i}\right)  }\sin\left(  \Omega
_{k}^{\left(  \lambda_{-}\right)  }t\right)  \cos\left(  \Omega_{k}^{\left(
\lambda_{+}\right)  }t\right)  ]\times\left[  \exp\left(  -2\beta\Omega
_{k}^{\left(  \lambda_{i}\right)  }\right)  -1\right] \nonumber\\
&  +2\exp\left(  -\beta\Omega_{k}^{\left(  \lambda_{i}\right)  }\right)  |,
\label{n21}%
\end{align}
where%
\begin{equation}
Z=\exp\left(  -2\beta\Omega_{k}^{\left(  \lambda_{i}\right)  }\right)
+1+2\exp\left(  -\beta\Omega_{k}^{\left(  \lambda_{i}\right)  }\right)
\label{n22}%
\end{equation}
is the partition function and $\beta$=$1/k_{B}T$ with $k_{B}$ the Boltzmann
constant. By changing $\beta$, we can get the evolution of the coherence
factor $F\left(  t\right)  $ at different temperatures. In Fig. \ref{fig4} we
show the numerical result of the coherence factor $F\left(  t\right)  $ as a
function of time $t$ and temperature $T$. The other parameters are
$\lambda_{i}$=$\lambda_{e}$=$1$, $\gamma$=$1$, $g$=$0.05$, and $N$%
=$2\times10^{2}$. It is within the expectation that the evolution of $F\left(
t\right)  $ for thermal equilibrium state at extremely high temperature drops
from unity to zero in a short time. Surprisingly, whereas, the decaying speed
of $F\left(  t\right)  $ dose not increase monotonously with temperature for
certain time $t$. In Fig. \ref{fig5} we show the cross-section of Fig
\ref{fig4} at different time $t$. One can see that there is a peak on every
line. That is, at certain temperature, the coherence factor $F\left(
t\right)  $ decays more slowly than that of the ground state. We think that
the decrease in the decay speed of $F\left(  t\right)  $ originates from the
increase in the proportion of the states $\left\vert 1\right\rangle
_{k}\left\vert 0\right\rangle _{-k}$ and $\left\vert 0\right\rangle
_{k}\left\vert 1\right\rangle _{-k}$ (on which there is no decoherence) in the
thermal equilibrium state. 

\begin{figure}[ptb]
\begin{center}
\includegraphics[width=0.8\linewidth]{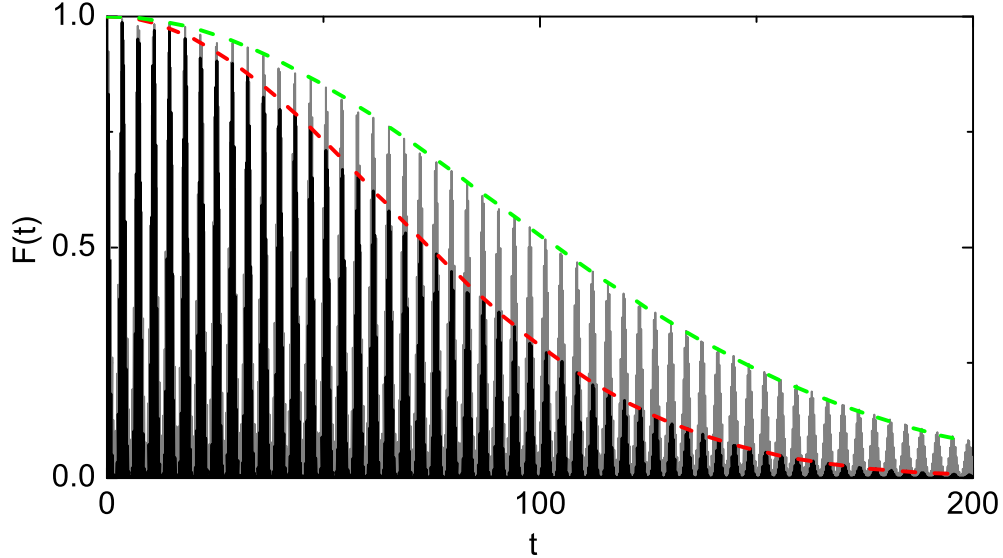}
\end{center}
\caption{(Color online). The coherence factor $F\left(  t\right)  $ as a
function of time $t$ for Ising ($\gamma$=$1$) spin chain with a size of $N$=$800$ in the
strong coupling regime $g$=$500$.}%
\label{fig6}%
\end{figure}

Now we turn to study the evolution of $F\left(  t\right)  $ with different
initial states in the strong coupling regime. In this regime, $F\left(
t\right)  $ will oscillate rapidly under a Gaussian envelope
\cite{Cucc2,Cormick2,Yuan2}. Following Refs. \cite{Cucc2,Cormick2}, we derive
a formula to approximate the Gaussian envelope. By noting that in the strong
coupling regime, $\alpha_{k}^{\left(  \lambda_{+-}\right)  }\approx\pi/2$, we
can simplify Eq. (\ref{n18}) as%
\begin{equation}
F\left(  t\right)  \approx\prod\nolimits_{k=1}^{M}|\exp\left[  it\left(
\Omega_{k}^{\left(  \lambda_{+}\right)  }+\Omega_{k}^{\left(  \lambda
_{-}\right)  }\right)  \right]  \cos^{2}\alpha_{k}^{\left(  \lambda
_{+i}\right)  }+\exp\left[  it\left(  -\Omega_{k}^{\left(  \lambda_{+}\right)
}-\Omega_{k}^{\left(  \lambda_{-}\right)  }\right)  \right]  \sin^{2}%
\alpha_{k}^{\left(  \lambda_{+i}\right)  }|. \label{n34}%
\end{equation}
The energy terms can be expressed as $\Omega_{k}^{\left(  \lambda_{+}\right)
}$+$\Omega_{k}^{\left(  \lambda_{-}\right)  }$=$E$+$\Delta_{k}$. The evolution
of $F_{k}\left(  t\right)  $ oscillates rapidly with almost the same frequency
$E\mathtt{\approx}4g$. The differences $\Delta_{k}\mathtt{\ll}E$ are
responsible for the decay of the envelope. By evaluation near the peaks of the
oscillations, $t$=$n\pi/E$+$\delta t$, and by using the Taylor expansions in
$\delta t$ and $\Delta_{k}$, we find that the frequency of the peaks
corresponds to the energy
\begin{equation}
E=\frac{\underset{k}{\overset{M}{%
{\displaystyle\sum}
}}\sin^{2}\left(  \theta_{k}^{g}-\theta_{k}^{i}\right)  \left[  \Omega
_{k}^{\left(  \lambda_{+}\right)  }+\Omega_{k}^{\left(  \lambda_{-}\right)
}\right]  }{\underset{k}{\overset{M}{%
{\displaystyle\sum}
}}\sin^{2}\left(  \theta_{k}^{g}-\theta_{k}^{i}\right)  }, \label{n35}%
\end{equation}
and the value of the envelope at these peaks can be approximated by
\begin{equation}
\tilde{F}\left(  t\right)  \approx\exp\left\{  -\tilde{s}_{M}^{2}%
t^{2}/2\right\}  , \label{n39}%
\end{equation}
where
\begin{equation}
\tilde{s}_{M}^{2}=\sum\left\{  \sin^{2}\left(  \theta_{k}^{g}-\theta_{k}%
^{i}\right)  \left\{  \left[  \Omega_{k}^{\left(  \lambda_{+}\right)  }%
+\Omega_{k}^{\left(  \lambda_{-}\right)  }\right]  -E\right\}  ^{2}\right\}  .
\label{n36}%
\end{equation}
By noting that for $\gamma$=$1$
\begin{align}
\underset{k}{%
{\displaystyle\sum}
}\sin^{2}\theta_{k}^{i}  &  \approx\left\{
\begin{array}
[c]{c}%
\frac{M}{2\lambda_{i}^{2}}\\
\frac{M}{2}%
\end{array}
\right.
\begin{array}
[c]{c}%
\lambda_{i}^{2}>1\\
\lambda_{i}^{2}\leq1
\end{array}
,\label{n37a}\\%
{\displaystyle\sum}
\left(  \sin^{2}\theta_{k}^{i}\sin^{2}\frac{2\pi k}{N}\right)   &
\approx\left\{
\begin{array}
[c]{c}%
\frac{M}{8}\frac{3\lambda_{i}^{2}-1}{\lambda_{i}^{4}}\\
\frac{M}{8}\left(  3-\lambda_{i}^{2}\right)
\end{array}
\right.
\begin{array}
[c]{c}%
\lambda_{i}^{2}>1\\
\lambda_{i}^{2}\leq1
\end{array}
,\label{n37b}\\%
{\displaystyle\sum}
\left(  \sin^{2}\theta_{k}^{i}\sin^{4}\frac{2\pi k}{N}\right)   &
\approx\left\{
\begin{array}
[c]{c}%
\frac{M}{32\lambda_{i}^{6}}\left(  10\lambda_{i}^{4}-5\lambda_{i}^{2}+1\right)
\\
\frac{M}{32}\left(  10-5\lambda_{i}^{2}+\lambda_{i}^{4}\right)
\end{array}
\right.
\begin{array}
[c]{c}%
\lambda_{i}^{2}>1\\
\lambda_{i}^{2}\leq1
\end{array}
, \label{n37c}%
\end{align}
and after a tedious calculation, we obtain the exact expression for $\tilde
{s}_{M}^{2}$ as%
\begin{equation}
\tilde{s}_{M}^{2}\approx\left\{
\begin{array}
[c]{c}%
\frac{M}{8g^{2}\lambda_{i}^{4}}\left(  \lambda_{i}^{2}+1\right) \\
\frac{M}{8g^{2}}\left(  \lambda_{i}^{2}+1\right)
\end{array}
\right.
\begin{array}
[c]{c}%
\lambda_{i}^{2}>1\\
\lambda_{i}^{2}\leq1
\end{array}
. \label{n38}%
\end{equation}
Similar to the case of weak coupling regime, here we firstly give the
expression for the reciprocal of the envelops's width $\tilde{s}_{M}^{2}$ with
the parameters of the Hamiltonian for Ising ($\gamma$=$1$) model. That is, we
succeed in using a simple formula to approximate the Gaussian envelope of the
evolution of the coherence factor. It is noteworthy that in the strong
coupling regime, the formula Eq. (\ref{n33}) is still applicable in very short
time ($t\ll1$). In Fig. \ref{fig6} we show the numerical results of $F(t)$ as
a function of $t$ with $\lambda_{i}$=$0.5$ and $\lambda_{i}$=$1.5$
respectively. The red and blue lines are drawn with Eqs. (\ref{n18}) and
(\ref{n39}) respectively with $\lambda_{i}$=$0.5$; The black and yellow lines
are drawn with Eqs. (\ref{n18}) and (\ref{n39}) respectively with $\lambda
_{i}$=$1.5$. And the value of $\tilde{s}_{M}^{2}$ in Fig. \ref{fig6} is
calculated with Eq. (\ref{n38}). One can see that the approximate envelope
fits very well. Also similar to the case of weak coupling regime, the
derivative of $\tilde{s}_{M}^{2}$ with $\lambda_{i}^{2}$ is not continuous at
the critical point $\lambda_{i}$=$1$.$0$ of the initial Hamiltonian.
In some previous papers \cite{Cucc1,Cormick1}, universal regime of
the Gaussian envelope has been found when the coupling strength is large
enough, which means that the envelope of the decay of the coherence factor in
the system is Gaussian with a width independent of the system-environment
coupling strength. While obviously, the width of the Gaussian envelope we
calculated is proportional to the coupling strength $g$. This seeming conflict
originates from the different settings of the Hamiltonian. In Refs.
\cite{Cucc1,Cormick1}, only one of $\lambda_{-}$ and $\lambda_{+}$ is
correlated with the coupling strength $g$, the other is uncorrelated with $g$.
This setting leads to different results of $F(t)$ and thus the width of the envelope.

In conclusion, we have studied the decaying process of the coherence factor of
a coupled system consisting of a central spin and its correlated environment
described by a general $XY$ spin-chain model. We mainly analyzed the
dependence of the decoherence on the initial state of the environmental
Hamiltonian by assuming that the initial state is the ground state of the
initial Hamiltonian $H_{E}^{\left(  \lambda_{i}\right)  }$ at time $t$=$0$
which may be different from the evolving Hamiltonian $H_{E}^{\left(
\lambda_{e}\right)  }$. In this case we have obtained the exact analytical
expression for the decoherence factor $F\left(  t\right)  $. At the critical
point $\lambda_{e}$=$1$ of the evolving Hamiltonian, the coherence factor
$F\left(  t\right)  $ decays rapidly from unity to zero for any value of
$\lambda_{i}$ and with no revivals when the site number $N$ is large enough.
The evolution of the coherence factor $F\left(  t\right)  $ as a function of
time $t$ is Gaussian in a short time whenever in weak or strong coupling
regime. In the strong coupling regime, $F\left(  t\right)  $ oscillates
rapidly under a Gaussian envelope. In this case, as a main result, we have
obtained a simple expression for the Gaussian decay and the Gaussian envelope
with the parameters of the Hamiltonian for Ising ($\gamma$=$1$) model. All
these approximate expressions fit the evolution of the coherence factor or the
envelope very well. We have also chosen thermal equilibrium state as initial
state and found that the decaying speed of $F\left(  t\right)  $ dose not
increase monotonously with temperature for certain time $t$.

This work was supported by NSFC under Grants No. 11147143, No. 11204012, and
No. 91321103.

\end{document}